\documentstyle[aps,prl,array,multicol]{revtex}
\begin{document}
\title{Glauber slow dynamics of the magnetization  
in a molecular Ising chain
}
\author{A. Caneschi,$^1$ D. Gatteschi,$^1$ N. Lalioti,$^1$ 
C. Sangregorio,$^1$ R. Sessoli,$^1$ G. Venturi,$^2$ 
A. Vindigni,$^2$ A. Rettori,$^{2,3}$
M. G. Pini,$^{4}$ and M. A. Novak$^5$}
\address{$^1$ 
Dipartimento di Chimica, Universit\`a
degli Studi di Firenze, Via Maragliano 75/77, 
50144 Firenze, Italy}
\address{$^2$ Dipartimento di Fisica,
Universit\`a degli Studi di Firenze,
Largo E. Fermi 2, I-50125 Firenze, Italy}
\address{$^3$ Istituto Nazionale per la Fisica della Materia,
Unit\`a di Firenze, Largo E. Fermi 2, I-50125 Firenze, Italy}
\address{$^4$ Istituto di Elettronica Quantistica, Consiglio Nazionale 
delle Ricerche, Via Panciatichi 56/30, I-50127 Firenze, Italy}
\address{$^5$
Instituto de Fisica, Universidade Federal do Rio de Janeiro, CP68528,
RJ 21945-970, Brazil}
\date{\today}
\maketitle
\begin{abstract}
The slow dynamics (10$^{-6}$ s - 10$^4$ s) of the magnetization 
in the paramagnetic phase, predicted by Glauber  
for 1d Ising ferromagnets, has been observed with
ac susceptibility   and SQUID magnetometry measurements in a molecular chain
comprising alternating Co$^{2+}$ spins and organic radical spins 
strongly antiferromagnetically  coupled. An Arrhenius behavior with activation
energy  $\Delta$=152 K has been observed for ten decades of relaxation time
and found to be consistent with the Glauber model. We have extended this model
to take into account the ferrimagnetic nature of the chain as well as its
helicoidal structure.

\end{abstract}

\begin{multicols}{2}
Slow magnetic relaxation in low dimensional systems 
without long range order is a very interesting subject 
both from the fundamental and the applicative point 
of view \cite{Science}.
This phenomenon has been clearly observed and 
theoretically explained in molecular magnetic clusters,  
which show also hysteretic effects of molecular 
origin \cite{Nature}. These clusters can be considered 
as magnetic quantum dots, i.e. zero-dimensional compounds.
In one-dimensional (1d) systems, 
a macroscopically slow relaxation was predicted 
many years ago by Glauber \cite{Glauber}  
for a ferromagnetic Ising chain.
In the last three decades many quasi-1d magnetic compounds
were synthesized and characterized and new features 
were observed, such as nonlinear excitations \cite{Mikeska},
Haldane gap \cite{Haldane} and spin-Peierls phase\cite{Peierls}.
 On the contrary, up to now, Glauber's prediction has not been 
experimentally confirmed because two strict conditions
must be met: i) a very strong easy axis anisotropy, 
ii) a very high ratio $J/J^{\prime}$, between the 
intrachain and the interchain exchange interactions, 
leading to a very low critical temperature 
to three-dimensional (3d) order. 

We have recently reported preliminary data 
which suggest that these conditions can be fulfilled 
in a real quasi-1d ferrimagnetic compound,  
[Co(hfac)$_2$NITPhOMe] (CoPhOMe in the following) \cite{Angew}. 
It has a trigonal symmetry with 
ternary helices \cite{Dalton}, along the $c$ axis, 
formed by Co(hfac)$_2$ moieties and by the NITPhOMe 
organic radicals, as outlined in Fig. 1.
The primitive cell contains three  Co$^{2+}$ metal ions
alternating with three radicals. Each Co$^{2+}$ is geometrically
related to the other two through a 
120$^{\circ}$ rotation around the $c$ axis and has a 
local axis of easy anisotropy, $z$, which makes an angle
$\theta \simeq 50^{\circ}$ with $c$, as observed in
monomeric complex of formula Co(hfac)$_2$(NITPhOMe)$_2$  
\cite{Preprint}, where the Cobalt ion coordination is practically the 
same as in CoPhOMe. The strong 1d character is 
a consequence of the strong metal-radical magnetic
exchange along the chain and of the absence of chemical bond 
between different chains. The temperature 
dependence of $\chi T$ gives a clear indication of
 1d behavior, with antiferromagnetic 
coupling between Co$^{2+}$ and NITPhOMe radical. Since 
the effective $S=1/2$ spin of the Co$^{2+}$ and the $S=1/2$ spin of the 
radical  are characterized by different $g$ values ($g_{Co}$ and $g_{R}$ 
in the following),  the chains behave as 1d ferrimagnets, due to the 
non-compensation of the magnetic moments.
An approximate quantitative analysis using the alternating 
spin Ising model suggested that the intrachain exchange interaction 
is of the order of 200 K in absolute value.
The ac magnetic susceptibility 
shows that the relaxation time of the magnetization below 
20 K follows a thermally activated
behavior. Below 4 K a stepped hysteresis was observed 
with the field parallel to the trigonal axis, while
no hysteresis was observed with the field perpendicular
to $c$. 

We wish to work out here an extension of the
Glauber dynamics to a helical Ising ferrimagnet, showing
how it can unambiguously justify the experimental features 
recalled in more detail here below.
In Fig. 2 the isothermal magnetization as a function 
of the field $H$ is reported at $T=2$ K for two 
different orientations of the field. When $H$ is applied 
along the $c$ axis, a hysteresis cycle  appears 
below 5 K, while the hysteresis is absent 
down to 2 K for fields applied perpendicularly to $c$. 
The unusual shape of the hysteresis (characterized by 
well defined steps at $H=\pm 4$ kOe)
is discussed later on. 
The magnetization, measured by cooling 
in zero field and then applying a field of 100 Oe,
shows the onset of irreversibility around 6 K, if
measured with the field parallel to the $c$ axis, 
while a paramagnetic behavior is observed down to 1.8 K
by applying the field in the trigonal plane.
The blocking of the magnetization at 6 K is 
not due to a phase transition to 3d order but 
has a dynamical nature, as confirmed by 
ac magnetic susceptibility measurements 
performed in the range 0.2 Hz - 95 kHz, which 
show a strong frequency dependence below 20 K: 
the in-phase susceptibility $\chi^{\prime}$ (not shown) goes 
through a maximum and the out-of-phase component 
$\chi^{\prime \prime}$ 
becomes different from zero and goes 
through a maximum (see Fig. 3), showing that the 
relaxation in this temperature range becomes slow in the 
time scale of the experiment (10$^{-6}$ s - 1 s).

The observed relaxation time $\tau$ follows an exponential 
law due to a thermally activated mechanism (see Fig. 4)
\begin{equation}
\tau=\tau_0~
\exp{\left\lbrack{{\Delta}\over {k_BT}}\right\rbrack}
\end{equation}
The best fit gives $\tau_0=4(1) \cdot  10^{-11}$ s, 
$\Delta/k_B=152(1)$ K. 
The value of $\tau_0$ is significantly smaller than 
that observed in superparamagnetic clusters \cite{Nature}\cite{Fe8EPL}. 
It is worth noticing 
 that $\chi^{\prime}$ goes to zero 
for $T$ below the peak in $\chi^{\prime \prime}$ 
and indeed $\chi^{\prime}$ and $\chi^{\prime \prime}$ 
almost describe a semicircle in the Cole-Cole 
plot (see the inset of Fig.3) \cite{Cole}.
The ac susceptibility recorded with the field oscillating 
in the trigonal plane is essentially paramagnetic except for
a small fraction, which we attribute to
the  non perfect alignment of the bunch of oriented crystals
used for these measurements.
For $T\le 5$ K the relaxation was also studied by 
monitoring the decay of the longitudinal magnetization 
which was found to be exponential, as a  
function of time, if weak fields ($<1$ kOe) were applied. 
The characteristic time $\tau$ is in agreement with the 
 law (1) and the parameters reported above.
The Arrhenius behavior is therefore obeyed in the
wide temperature range 4.5 K - 15 K and 
over ten decades in time (see Fig. 4).
In a real system interchain interactions are always 
present, however the 3d ordering cannot be detected
if the critical temperature is 
well below the magnetization freezing temperature. 
This  hypothesis is supported by the lack of 
anomalies in the specific heat \cite{Novak} and in 
preliminary neutron diffraction spectra \cite{Radaelli}.

The dynamic behavior is therefore strongly reminiscent of 
that observed in molecular clusters \cite{Nature} \cite{Fe8EPL}, where the 
energy barrier which hampers the reversal of
the magnetization is originated by the magnetic 
anisotropy. In the present case, despite the 
large value of the barrier, the plane orthogonal
to the trigonal axis is not a hard plane, as the 
magnetization rapidly approaches saturation 
(see Fig. 2). This anomaly appears to be related
to the helicoidal structure of the chain, as 
demostrated by the following extension of the
Glauber theory to a helicoidal ferrimagnetic chain.

According to the Glauber model \cite{Glauber}, 
 the \( j \)-th spin \( \sigma _{j} \) is assumed to take the integer values
\( \pm 1 \), corresponding to two possible projections along
its local \( z \) axis. The \( \sigma _{j} \) coordinates depend stochastically
on time. If we consider an isolated spin, the probability per unit time to
reverse is given by \( \frac{1}{2}\alpha _{0} \), where \( \alpha _{0} \) is a
free parameter of the model. In the case of \( N \) coupled spins we
need another phenomenological parameter, \( \gamma  \), which accounts for the
tendency of the \( j \)-th spin to align parallel to its nearest neighbors:
i.e., the transition probability from the state \( \sigma _{j} \) to the state
\( -\sigma _{j} \) is assumed to be \cite{Glauber}: 
\begin{equation}
w_{\sigma _{j}\rightarrow -\sigma _{j}}=
{1\over 2}\alpha _{0}[1-{1\over2 }\gamma 
\sigma _{j}(\sigma _{j+1}+\sigma _{j-1})]
\end{equation} 
 Imposing the detailed balance it is possible to obtain a correspondence
between \( \gamma  \) and the exchange constant \( J \) of the Ising model
Hamiltonian.
When the system has reached equilibrium at
temperature \( T \), the ratio of the transition probabilities \( w_{\sigma
_{j}\rightarrow -\sigma _{j}} \) and \( w_{-\sigma _{j}\rightarrow \sigma
_{j}} \) $must$ equal the ratio of the Maxwell-Boltzmann probabilities
associated with the two equilibrium configurations \( (\sigma _{1},...,\,
-\sigma _{j},...,\, \sigma _{N}) \) and  \( (\sigma
_{1},...,\, \sigma _{j},...,\, \sigma _{N}) \) \cite{Glauber}. This yields \(
\gamma ={\textrm{tanh}}(J/2k_{B}T) \).

In the Glauber model the direction of each local
\( z \) axis is determined by the applied field 
while in our case we will consider
as \( z \) axis of each Co$^{2+}$ the local axis of easy anisotropy. So
inside the same cell we will have three different \( z \) axes for the  
Co$^{2+
}$  ions (\( z_{2} \), \( z_{4} \), \( z_{6} \)) and three for the
radicals (\( z_{1} \), \( z_{3} \), \( z_{5} \), which are given by the
vectorial sum of the directions of magnetic anisotropy of the two  Co$^{2+}$
nearest neighbours). The extended Ising 
hamiltonian of this system of spins can be written:
\begin{eqnarray} 
&{\cal H}=-\sum_{n=1}^{N}\sum_{m=1}^{3} 
\Bigg\{ 
{1\over4}J\Bigl( 
\sigma_{n,2m-1} \sigma_{n,2m} 
\Bigr)+ 
\cr
&+{1\over 2}\mu_B   
\Bigl( g_{R}\sigma _{n,2m-1}\underline{H}\cdot \underline{e}_{2m-1}
+g_{Co}\sigma _{n,2m}\underline{H}\cdot \underline{e}_{2m}
\Bigr) 
\Bigg\} 
\end{eqnarray} 
where \( \underline{e}_{j} \) is the unitary vector corresponding to the \(
z_{j} \) direction. In the coordinate $\sigma_{n,j}$,  $n$ represents the
cell index and  $j$ the magnetic center inside the $n$-th cell.
Since we are interested in the ac response of the sample to an applied
oscillating field, we can focus on the stationary solution of the problem.
So the system of \( 6N \) coupled equations,
describing the Markov relaxation process of the spins chain, reduces to six
coupled differential equations for the expectation values \( z_{j}(t)=\langle
\sigma _{n,j}(t)\rangle  \) (\( j=1,\cdots ,6 \)). In fact
the translational invariance allows  to neglect the $n$ index. 
The equation describing the evolution of the
system can be written in matricial form as \begin{equation} {{\partial
\vec{z}}\over {\partial (\alpha _{0}t)}}=A\vec{z}+\vec{k}e^{-i\omega t}
\end{equation} The \( \vec{k} \) vector depends on the orientation of the
oscillating magnetic field with respect to the local axis of each spin and its
elements will be specified later on. The symmetric matrix \( A \) has elements
\( A_{jj}=-1 \), \( A_{j,j+1}=A_{j-1,j}=A_{1,6}=A_{6,1}=\gamma /2 \) and zero
otherwise.

\noindent Denoting by \( U \) the orthogonal matrix which diagonalizes \( A \),
\( D=U^{\dagger}AU \), we perform the transformations \( {\vec{\zeta
}}=U^{\dagger}\vec{z} \), \( \vec{\kappa }=U^{\dagger}\vec{k} \), so
that the system (4) is decoupled as \begin{equation}
{{\partial \vec{\zeta }}\over {\partial (\alpha _{0}t)}}=
D\vec{\zeta }+\vec{\kappa }e^{-i\omega t}
\end{equation}
 where the diagonal matrix \( D \) has elements 
\( D_{11}=-1+\gamma  \), \( D_{22}=-1-\gamma  \),
\( D_{33}=D_{44}=-1+\gamma /2 \), \( D_{55}=D_{66}=-1-\gamma /2 \). 
Each equation of the decoupled system (5) 
can be integrated, in the stationary limit, to
give  \begin{equation} \zeta _{j}(t)=\kappa _{j}{{\alpha _{0}}\over
{\alpha _{j}-i\omega }}e^{-i\omega t} \end{equation}
 with \( \alpha _{1}=(1-\gamma )\alpha _{0} \); 
\( \alpha _{2}=(1+\gamma )\alpha _{0} \);
\( \alpha _{3}=\alpha _{4}=(1-\gamma /2)\alpha _{0} \); 
\( \alpha _{5}=\alpha _{6}=(1+\gamma /2)\alpha _{0} \).
Since \( |\gamma |<1 \), we have that \( \alpha _{3} \), \( \alpha _{4} \),
\( \alpha _{5} \), \( \alpha _{6} \) cannot vanish. On the contrary, for 
\( T\rightarrow 0 \),
\( \alpha _{1} \) and \( \alpha _{2} \) can vanish for \( J>0 \) and \( J<0 \)
respectively. In our case \( J<0 \) and the slow relaxation (exponentially
diverging relaxation time) is expected for \( |J|/k_{B}T\gg 1 \) due to the
mode \( \zeta _{2} \)\begin{equation}
\tau _{2}={1\over {\alpha _{2}}}={{e^{-J/k_{B}T}}\over {2\alpha _{0}}}
\end{equation}
 while \( \tau _{1}\simeq 1/\alpha _{0} \). 
The law (7) is in agreement with the  
fit (1) where  $\Delta=-J$ and  
$\tau_{0}$=$1\over{2\alpha_0}$.  

 When the magnetic field is applied along the $c$ axis, its
projection along each local axis \( \underline{e}_{i} \) is trivially \(
H(t)\cos \theta  \) and, while \( \kappa _{3\Vert }=\kappa _{4\Vert }=\kappa
_{5\Vert }=\kappa _{6\Vert }=0 \), one has \begin{equation}
\kappa _{2\Vert }={3\over {\sqrt{6}}}{{\mu _{B}H_{0}S}\over 
{k_{B}T}}{{1-\eta ^{2}}\over {1+\eta ^{2}}}\cos \theta (g_{Co}-g_{R})
\end{equation}
where \( \eta =\tanh \left( \frac{J}{4k_{B}T}\right)  \). 
For \( \kappa _{1\Vert } \)
the difference between the Land\'{e} factors is substituted by the addition.
The complex susceptibility can be obtained through the projection of the
contribution coming from each spin: 
\begin{eqnarray} 
\chi\left(\omega\right)=S^2 \mu_{B}^{2} \cos^{2}\theta 
\frac{N}{k_B T}
\frac{1-\eta^2}{1+\eta^2}  
\cr
\Big\{
\alpha_{0}\frac{\left(g_{Co}+g_{R}\right)^2}{\alpha_{1}-i\omega}+ 
\alpha_{0}\frac{\left(g_{Co}-g_{R}\right)^2}{\alpha_{2}-i\omega}
\Big\} 
\end{eqnarray} 
where, since in the experimental conditions  (frequency and temperature)
 $\omega \ll \alpha_1$, the first
term is negligible.
Hence, by varying the frequency,  $\chi^{\prime \prime}$ as a function
of $\chi^{\prime}$ is expected to describe a semicircle,
 in agreement with the experimental data reported in Fig. 3.  If the
oscillating magnetic field is applied along a generic perpendicular direction
with respect to the $c$ axis, after a few steps we find \( \kappa _{1\perp
}=\kappa _{2\perp }=0 \) and consequently the slow mode \( \zeta _{2} \) is
not activated but only a linear combination of \( \zeta _{3} \), \( \zeta _{4}
\), \( \zeta _{5} \), \( \zeta _{6} \) is.   These results explain the
experimental evidence for  slow relaxation for magnetic fields applied
parallel to  the $c$ axis  while, for magnetic fields perpendicular to $c$,
the relaxation time  
$\tau$ remains of the  order of $\frac{1}{\alpha_{0}}$. 

The hysteresis curves reported in Fig. 2 are characterized 
by well-defined steps, the more evident observed at $H=\pm 4$ kOe.
The steps observed in the first and third quadrants, i.e. when 
the magnetization does not change sign, have a static nature,
as confirmed by the fact that they are fully reversible 
(see inset of Fig. 2),
do not depend on the sweeping rate of the field, and
are observable above the blocking temperature, when
the hysteresis disappears. Therefore, their origin seems
to be different from the resonant tunneling mechanism observed 
in molecular magnetic clusters \cite{Nature2} \cite{Clusters}.
Similar steps of static nature were observed
in linear antiferromagnetic chains with 
alternating magnetic moments when the Zeeman energy is comparable
to the exchange interaction \cite{Alternate}. In the present case
we expect the exchange interaction to be much
larger than the Zeeman energy in a field of 4 kOe.
However  the helical  structure and the fact that 
the local $z$ axes are not pointing along the axis of the helix
($\theta \simeq 50^{\circ}$)  can account for this anomaly as well as  
for the similar saturation values of the two components of the
magnetization (Fig. 2) despite the completely different dynamic behavior.
This hypothesis has been confirmed by classical transfer  
matrix calculations
and by diagonalization of a finite   quantum ring with six spins, 
where the trigonal symmetry of the helical structure  
has been taken into account. 
A detailed description of this study will be 
the subject of a future publication.

In  conclusion, we have shown that CoPhOMe is a 
quasi-1d ferrimagnetic 
compound which at low temperature shows  
slow dynamics  of the macroscopic magnetization 
of the type predicted many
years  ago by Glauber and never observed before.
The presence of magnetic hysteresis in the
absence of 3d magnetic order opens the exciting
possibility of storing information in an array
of magnetic centers with a dramatic reduction
in the dimensions (two of the three) of the
magnetic memory unit. From a more fundamental
point of view, we have extended the theory developed 
by Glauber for the simple model of an 
Ising ferromagnetic chain 
to the case of helicoidal geometry, and the anomalous 
dynamic behavior of CoPhOMe turned out to be 
completely understandable in terms of it.

We thank P. Radaelli and M. Affronte for the preliminary
informations on the neutron scattering and the specific heat 
data respectively, as well as W. Wernsdorfer for stimulating 
discussions. The financial support of Italian MURST and CNR, 
of the  EC through the TMR program 3MD (n$^{o}$ ERB4061PL97-0197) 
and MOLNANOMAG (n$^{o}$ HPRN-CT-1999-0012)
and  of Brasilian CNPq and FUJB is  acknowledged.

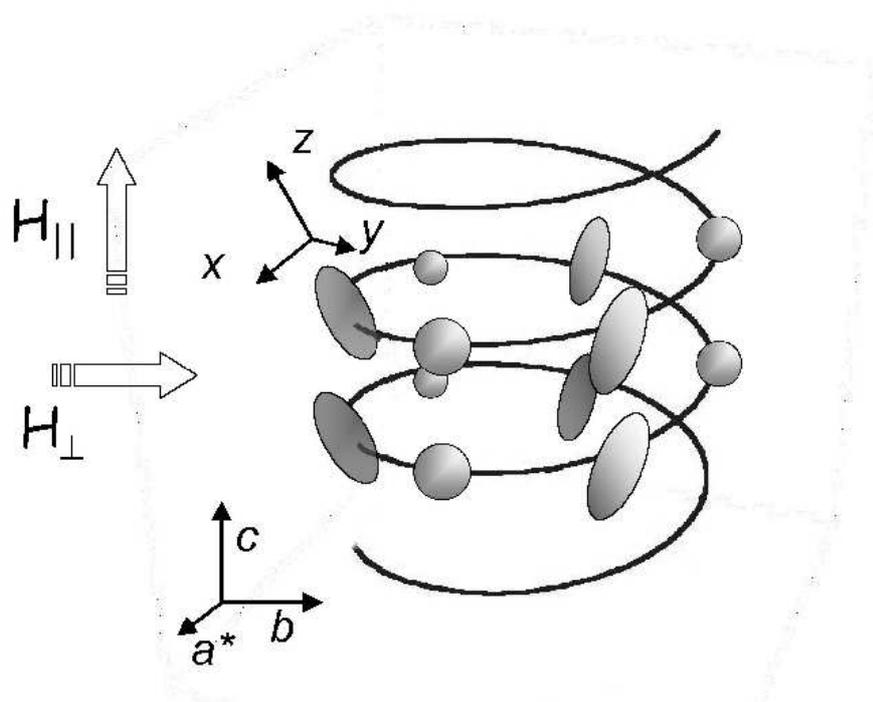
\begin{figure}
\caption{
Schematic view of the helicoidal structure of CoPhOMe.
The laboratory reference frame corresponding to the crystallographic
axes $a^*$, $b$, $c$ and the local reference frame 
($x$, $y$, $z$) are shown. The angle between $z$ and the $c$
axis is the $\theta$ angle mentioned in the text.
}
\end{figure}
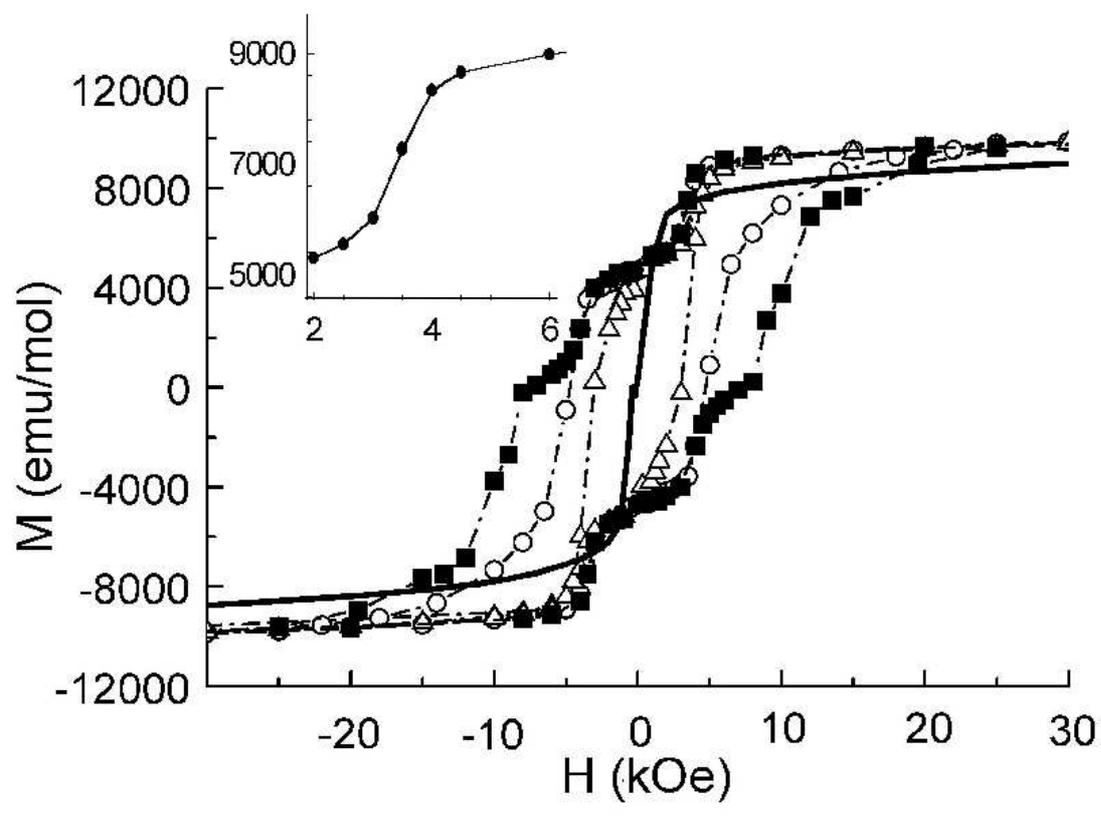
\begin{figure}
\caption{
Hysteresis loops recorded on a single crystal of CoPhOMe by
applying the field in the $a^*b$ plane (solid line) at $T=2$ K
and along the $c$ axis at 2.0 K (black squares), 3.5 K (open
circles) and 4.5 K (open triangles). In the inset: at $T=2$ K
the field was cycled three times around the step at 4 kOe.
The curves taken with increasing field (solid circles) and
decreasing field (solid line) are superimposable, denoting
the reversible character of the step.
}
\end{figure}
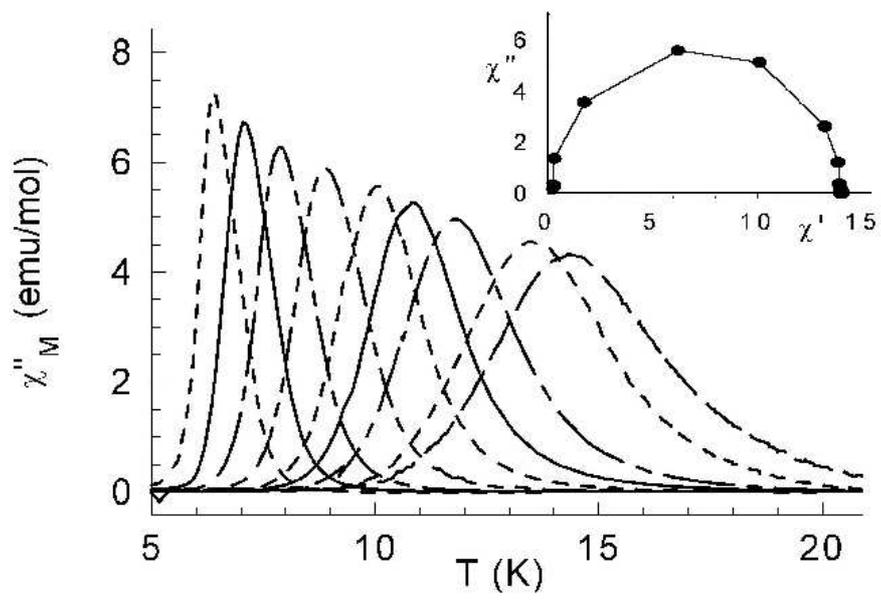
\begin{figure}
\caption{
Temperature dependence of the imaginary component of the
ac magnetic susceptibility measured in zero static field
and with the dynamic field oscillating parallel to the $c$ 
axis for nine selected frequencies in the
range 0.18 Hz - 95 kHz. In the inset 
$\chi^{\prime\prime}\,vs\, 
\chi^{\prime}$ (i.e, the Cole-Cole plot)
measured at 10 K.
}
\end{figure}
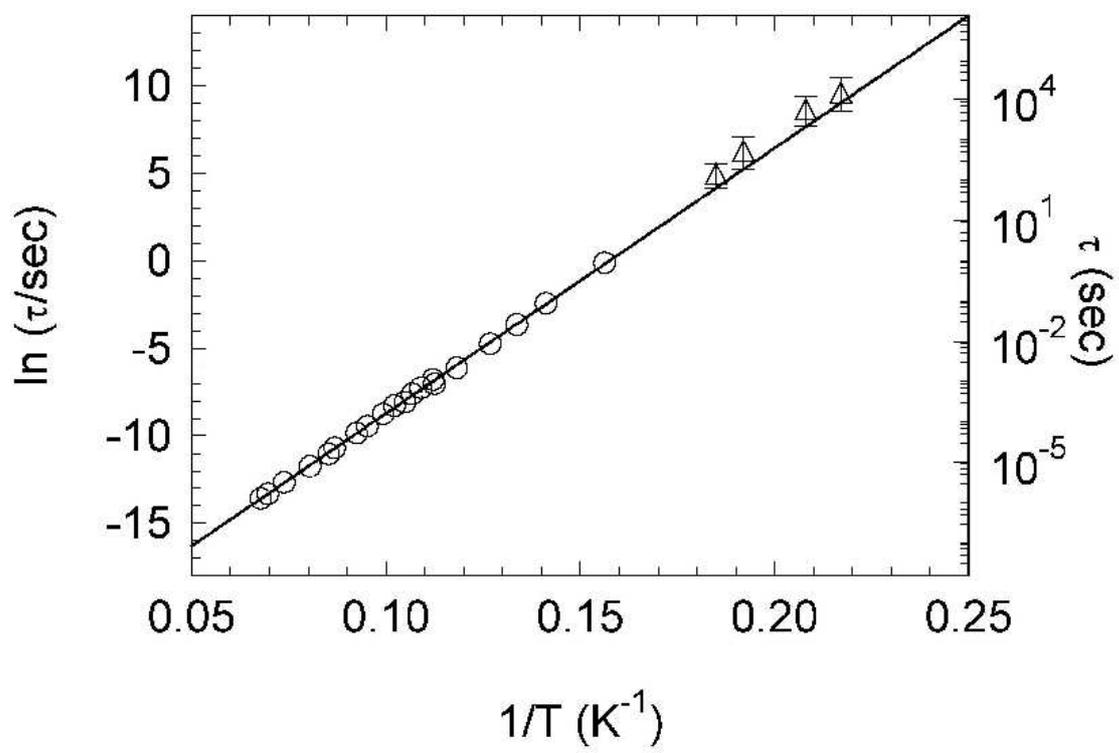
\begin{figure}
\caption{
Temperature dependence of the relaxation time.
The data represented by circles have been obtained
from the curves of Fig. 3 by assuming that, at the 
temperature of the maximum in each curve, $\tau=
1/(2\pi\nu)$, where $\nu$ is the frequency of
the oscillating field. The triangles represent
the relaxation time obtained from the
decay of the magnetization after having applied 
and removed a weak field.
The solid line represents the fit of the 
$\chi^{\prime\prime}$ ac data using Eq. (1). The best fit
parameters are $\Delta=152(\pm 1)$ K and 
$\tau_0=4(\pm 1) 10^{-11}$ s.
}
\end{figure}
\end{multicols}
\end{document}